\documentclass{iau} 
\usepackage{graphicx}

\title[Sun-as-a-star Helioseismic Observations] 
{Un-interrupted Sun-as-a-star Helioseismic Observations over Multiple Solar Cycles}

\author[Jain et al. ] 
{Kiran Jain$^1$, Sushanta Tripathy$^1$, Frank Hill$^1$, 
David Salabert$^{2,3}$, \\
Rafael A. Garc{\'i}a $^{2,3}$ and Anne-Marie Broomhall$^{4,5}$ }

\affiliation{$^1$National Solar Observatory, 3665 Discovery Drive, Boulder, CO 80303, USA\\ 
email: {\tt kjain@nso.edu}, {\tt stripathy@nso.edu}, {\tt fhill@nso.edu} \\[\affilskip]
$^2$IRFU, CEA, Université Paris-Saclay, F-91191 Gif-sur-Yvette, France\\
$^3$Laboratoire AIM, CEA/DRF-CNRS-Université Paris Diderot,F-91191 Gif-sur-Yvette, France\\
$^4$Department of Physics, University of Warwick, Coventry, CV4 7AL, UK\\
$^5$Institute of Advanced Study, University of Warwick, Coventry, CV4 7HS, UK 
}

\pubyear{2018}
\volume{340}  
\jname{Long-Term Datasets for the Understanding of Solar and Stellar Magnetic Cycle}
\editors{eds.}
\begin{document}

\maketitle

\begin{abstract}
 We analyze Sun-as-a-star observations spanning over solar cycles 22 -- 24  
from the ground-based network BiSON and solar cycles 
23 -- 24 collected by the space-based  VIRGO and GOLF instruments 
on board the {\it SoHO} satellite.  Using simultaneous 
observations from all three instruments, our analysis suggests that the structural and magnetic changes responsible 
for modifying the frequencies remained comparable between cycle 23 and cycle 24 but differ
from cycle 22. Thus we infer that the magnetic layer of the Sun has become thinner 
since the beginning of cycle 23 and continues during the current cycle. 
\end{abstract}

\firstsection 
\section{Introduction}

Long-term observations of solar activity clearly show the variable nature of Sun's magnetism.
These observations, e.g. in the form of sunspot counts and groups, exist for the past several centuries and 
provide large databases to understand the long-term as well as short-term trends in solar 
variability. In addition, a database of cosmogenic isotope variations is also available  for a much longer 
period providing long-term records of solar geomagnetic activity. However, such records to understand
the variations below the surface are only a few decades old and consist of the inferences obtained 
from the resonant $p$-mode oscillations. The frequencies of oscillating modes are modified by the 
mechanical properties of the layers through which the waves traverse, and provide insights for the 
study of structural and dynamical changes occurring in the Sun's interior. 
These frequencies vary with the changing level of magnetic activity and display
strong correlations with the measures of solar activity (e.g., 
\cite[Broomhall \& Nakariakov 2015]{Broomhall15}, \cite[Jain et al. 2009]{Jainetal09},
\cite[Salabert et al. 2015]{Salabert etal15}, and references therein).
Various oscillation data sets  are now available  which cover a wide range of  modes and are
sensitive to different layers of the interior. In this paper, however, we will only discuss 
low-degree modes that penetrate the solar core. It may be noted that the low-degree modes 
are generally better determined from the Sun-as-a-star observations which implies that the 
observables e.g. velocity, intensity etc.  are integrated over the entire disk. 

\section{Un-interrupted Sun-as-a-star Observations}

Since the discovery of solar oscillations, there have been several instruments which provide
data for the Sun-as-a-star observations. In this study, we use following data sets
which cover the period of more than a solar cycle.

{\bf Birmingham Solar Oscillation Network (BiSON)}: The BiSON instruments use the technique 
of resonance scattering spectroscopy to make Doppler observations in the K line at 769.9 nm. 
The network has been operational since 1981 and its six-site configuration was completed in 1992
(\cite[Davies et al. 2014]{Davies etal14}, \cite[Hale et al. 2016]{Hale etal16}).

{\bf Global Oscillations at Low Frequencies (GOLF)}:
GOLF  onboard {\it SoHO} measures the radial velocity Doppler shift, integrated
over the solar surface, in the Na D1 and D2 Fraunhofer lines at 589.6 
and 589.0 nm, respectively (\cite[Gabriel et al. 1995]{Gabriel etal95},
\cite[Garc{\'i}a et al. 2005]{Gracia etal05}).
It has been providing high-quality data since April 1996 
covering the solar activity cycles 23 and 24 with two short gaps in 1998-99 
due to the temporary loss  of  the  {\it SoHO} spacecraft.

{\bf Variability of Irradiance and Gravity Oscillations (VIRGO)}: 
 VIRGO  onboard {\it SoHO} is composed of three Sun photometers
(SPM) at 402 nm (Blue), 500 nm (Green) and 862 nm (Red) and provides
integrated intensities in these three wavelengths 
(\cite[Fr{\"o}hlich et al. 1996]{Frohlich etal95}). The coverage of VIRGO data is 
similar to GOLF.

\begin{figure}[t]
\begin{center}{
 \includegraphics[width=3.in]{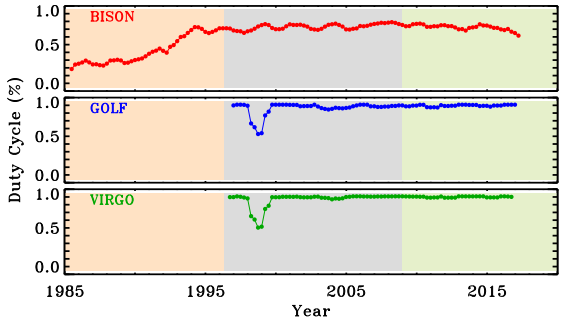}
  }
 \caption{Duty cycle of the data sets used in this study.} 
   \label{fig1}
\end{center}
\end{figure}

\section{Analysis and results}

We analyzed  32 years of continuous observations starting from Jan 1, 1985 from the BiSON network and 21 years 
of observations from the GOLF  and VIRGO instruments. The frequency tables are computed for 365 day subsets 
that overlapped by 91.25 days. It is well known that long-term data often suffer from 
the degradation of instruments and/or other observing environments. Thus, to minimize the bias, 
the start and end times in the time series have been selected so that where contemporaneous data
exist the subsets are temporally aligned.  The BiSON data span over three solar cycles (22 -- 24) 
and constitute 128 data sets while GOLF and VIRGO data cover cycles 23 -- 24  and constitute 81 and 82 
data sets, respectively. Duty cycles for all instruments are plotted in  Figure \ref{fig1}. 
It is observed that the duty cycles for GOLF and VIRGO are above 96\% with a strong dip during
the loss of {\it SoHO}.  However, in case of BiSON, the duty cycle increased with time due to the 
deployment of more sites, reached a maximum value of 84\% and remained consistent after the
network was completed.  
 
The frequency shifts, $\delta\nu$, were calculated for  $\ell$= 0 -- 3 modes in the frequency range
of 1860 $\mu$Hz $\le \nu <$ 4250  $\mu$Hz. Using the BiSON data until the rising phase of 
cycle 24,  \cite[Basu et al. (2012)]{Basu etal12} suggested that the changes in the Sun affecting 
the oscillation frequencies in cycle 23 were localized mainly to layer above about 0.996$R_{Sun}$.
Following their suggestion and the evidence that different frequencies have maximum sensitivity in 
different layers near the surface, we further carried out our analysis in four different frequency 
ranges. These  ranges are:  low (1860 $\le \nu <$ 2400 $\mu$Hz), mid (2400 $\le \nu <$ 2920 $\mu$Hz), high 
(2920 $\le \nu <$ 3450 $\mu$Hz) and very high (3450 $\le \nu <$ 4250 $\mu$Hz). While modes in the
 low frequency range have maximum sensitive to 0.996$R_{Sun}$, those in the high frequency range 
 are sensitive to  0.999$R_{Sun}$.

 \begin{figure}
\begin{center}
{
 
   \includegraphics[width=3.7in,angle=90]{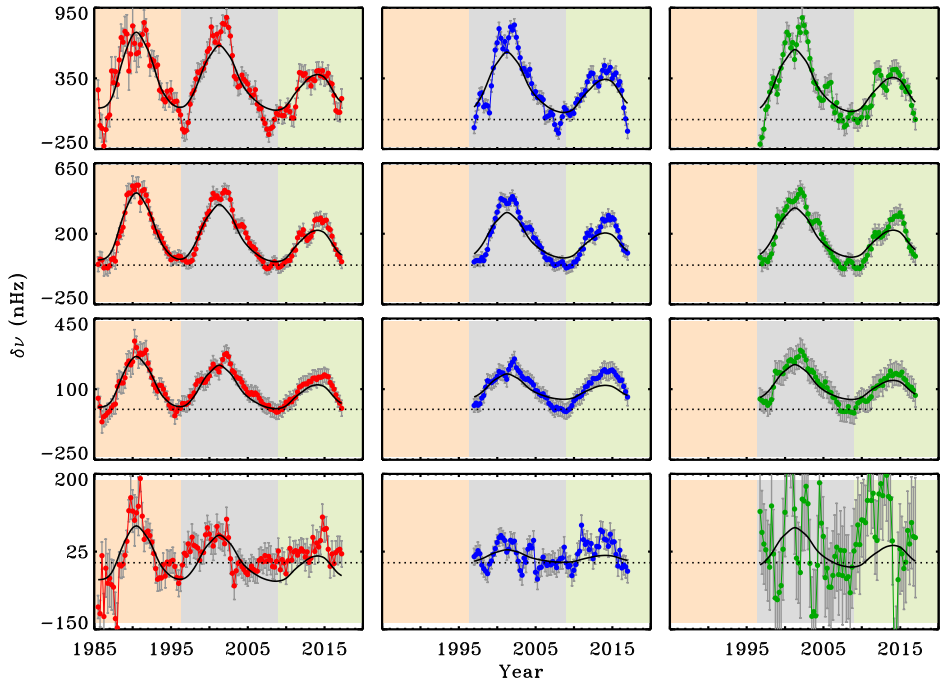} 
 }
  \caption{Average frequency shifts observed by BiSON (left), GOLF (middle) and VIRGO (right) as a function of time for four 
  frequency ranges; Top: 3450 $\le \nu <$ 4250 $\mu$Hz, 2nd from the top: 2920 $\le \nu <$ 3450 $\mu$Hz,
  3rd from the top: 2400 $\le \nu <$ 2920 $\mu$Hz, and bottom: 1860 $\le \nu <$ 2400 $\mu$Hz. The symbols in each panel show
the average shift in frequency while the uncertainties are shown by grey. The solid black line depicts
the scaled 10.7 cm radio flux variation. }
   \label{fig2}
\end{center}
\end{figure}

 We display the temporal variation of $\delta\nu$ in all four frequency bands in Figure \ref{fig2}. It is 
 evident that the VIRGO shifts in the low frequency range are noisier than the other two instruments. We also note
 that the uncertainties in the BiSON shifts in the rising phase of cycle 22 are significantly higher probably 
 due to the low duty cycle as shown in Figure~\ref{fig1}. In addition, the uncertainties in the shifts obtained 
 from modes in the highest frequency range is larger than mid and high frequency ranges indicating that the 
 goodness of fit reduces with the increase of frequencies after certain frequency.This is because the widths
 of the peaks in the power spectrum increases so substantially that it is difficult to constrain the frequency. 
 Furthermore, the amplitude (or S/N) is lower here.

The scaled variation of F10.7 cm radio flux (\cite[Tapping 2013]{Tapping13}) averaged over the same epochs 
that were used to calculate frequency shifts is over plotted in Figure~\ref{fig2}. The correlation between 
frequency shifts and radio flux is tested 
by calculating the Pearson's linear correlation (see Table \ref{tab1}). Comparing correlation coefficients
between cycles 22, 23 and 24 for BiSON, we find that there is a small change in mid and high frequency ranges
while it decreased significantly in cycles 23 and 24 in low frequency range. This implies that the layer sensed
by the low-frequency  modes
has changed with time and this change occurred either during the declining phase of cycle 22 or later. It 
should be further noted
that the correlation for the low-frequency modes  is about 30 to 40 percent lower than the other frequency 
ranges. In order to precisely determine the period of this change, a detailed phasewise correlation is 
required and will be presented in a future publication. Unfortunately, GOLF and VIRGO instruments became
operational at the beginning of cycle 23 and cannot be used to confirm the results obtained from the BiSON 
data for cycle 22. However, the agreement between results for cycle 23 and 24 from all the three instruments 
confirm that the Sun has gone through the near-surface structural changes in last few solar cycles and continues 
during the present cycle.  
 
\begin{table}[t]
  \begin{center}
  \caption{Correlation statistics between frequency shifts and 10.7 cm radio flux.}
  \label{tab1}
 {\scriptsize
  \begin{tabular}{|c|c|c|c|c|c|c|c|c|c|c|c|c|c|c|c|c|c|}\hline 
  &\multicolumn{4}{c} {\bf BiSON} && \multicolumn{4}{c} {\bf GOLF} &&\multicolumn{4}{c} {\bf VIRGO} & \\ \hline 
 {\bf Frequency}  &\multicolumn{4}{c} {\bf Solar Cycle} && \multicolumn{4}{c} {\bf Solar Cycle} &&\multicolumn{4}{c} {\bf Solar Cycle}&  \\ 
   {\bf Range}($\mu$Hz)              &22  & 23 & 24   & 22--23   & 23--24   &  22  & 23 & 24   & 22--23   & 23--24    &22  & 23 & 24   & 22--23   & 23--24\\ \hline
  1860 $\le \nu <$ 2400   & 0.79 & 0.58 & 0.63 & 0.66  & 0.56&   - & 0.59 &  0.65 &  -   & 0.54   &  -       & 0.41 &  0.19   &-  &0.59 \\ 
  2400 $\le \nu <$ 2920&  0.95 & 0.96 & 0.96 &0.94  & 0.95   &   - & 0.94 &  0.95 &  -   & 0.95   &  -     & 0.95    &0.94  &-  &  0.94 \\ 
  2920 $\le \nu <$ 3450&  0.97 & 0.97 & 0.97 &  0.97  & 0.97   & - & 0.96 &  0.97 &  -   & 0.96   &  -    & 0.96    & 0.97   &- &  0.96 \\ 
  3450 $\le \nu <$ 4250&  0.87 & 0.96 & 0.95 &   0.92  & 0.86   &  - & 0.95 &  0.92 &  -    & 0.94    &-   & 0.95   & 0.94   &- &  0.94 \\ \hline
  \end{tabular}
  }
 \end{center}
\end{table}

\section{Summary}
Long-term simultaneous Sun-as-a-star observations from three instruments illustrate  
that  structural and magnetic changes responsible 
for the frequency shifts remained comparable between cycle 23 and cycle 24, and are different
from cycle 22. We have also demonstrated that the magnetic layer of the Sun has become thinner 
in last two solar cycles.  However,  these low-degree 
modes also penetrate deeper however and there are 
differences when the modes confined to  the convection zone are analyzed. 
For example, the minimum sensed by these modes happened around the same time as in the solar activity 
indicators while the low-degree modes sensed minimum about a year earlier 
(\cite[Jain et al. 2011]{Jainetal11}, \cite[Salabert et al. 2009]{Salabert etal09},
\cite [Tripathy et al. 2010]{Tripathytal10}). Thus, these findings 
advocate the need of a long-term data base of solar oscillations (similar to solar activity) 
in order to understand the variability of different layers in the solar interior and its link to 
the surface magnetic activity.

SoHO is a project of international collaboration between ESA and NASA.
This work also utilizes  data  collected  by  the
Birmingham  Solar  Oscillations  Network  (BiSON),  which  is
funded by the UK Science Technology and Facilities Council
(STFC).



\end{document}